\documentclass[]{pasj01_revised}

\newcommand{\dd}{\mathrm{d}}

\begin{document}

\title{SKA-Japan Pulsar Science with the Square Kilometre Array}

\author{Keitaro \textsc{Takahashi}\altaffilmark{1},
Takahiro \textsc{Aoki}\altaffilmark{2},
Kengo \textsc{Iwata}\altaffilmark{3},
Osamu \textsc{Kameya}\altaffilmark{4},
Hiroki \textsc{Kumamoto}\altaffilmark{1},
Sachiko \textsc{Kuroyanagi}\altaffilmark{3},
Ryo \textsc{Mikami}\altaffilmark{5},
Atsushi \textsc{Naruko}\altaffilmark{6},
Hiroshi \textsc{Ohno}\altaffilmark{7},
Shinpei \textsc{Shibata}\altaffilmark{8},
Toshio \textsc{Terasawa}\altaffilmark{5},
Naoyuki \textsc{Yonemaru}\altaffilmark{1},
Chulmoon \textsc{Yoo}\altaffilmark{3}
(SKA-Japan Pulsar Science Working Group)
}
\altaffiltext{1}{Kumamoto University, Japan}
\altaffiltext{2}{Waseda University, Japan}
\altaffiltext{3}{Nagoya University, Japan}
\altaffiltext{4}{National Astronomical Observatory of Japan, Japan}
\altaffiltext{5}{Institute for Cosmic Ray Research, University of Tokyo, Japan}
\altaffiltext{6}{Tokyo Institute of Technology, Japan}
\altaffiltext{7}{Tohoku Bunkyo College, Japan}
\altaffiltext{8}{Yamagata University, Japan}
\email{keitaro@sci.kumamoto-u.ac.jp}


\maketitle

\begin{abstract}
The Square Kilometre Array will revolutionize pulsar studies with its wide field-of-view, wide-band observation and high sensitivity, increasing the number of observable pulsars by more than an order of magnitude. Pulsars are of interest not only for the study of neutron stars themselves but for their usage as tools for probing fundamental physics such as general relativity, gravitational waves and nuclear interaction. In this article, we summarize the activity and interests of SKA-Japan Pulsar Science Working Group, focusing on an investigation of modified gravity theory with the supermassive black hole in the Galactic Centre, gravitational-wave detection from cosmic strings and binary supermassive black holes, a study of the physical state of plasma close to pulsars using giant radio pulses and determination of magnetic field structure of Galaxy with pulsar pairs.
\end{abstract}

\begin{figure}
\vspace{-188mm}
\includegraphics[width=25mm]{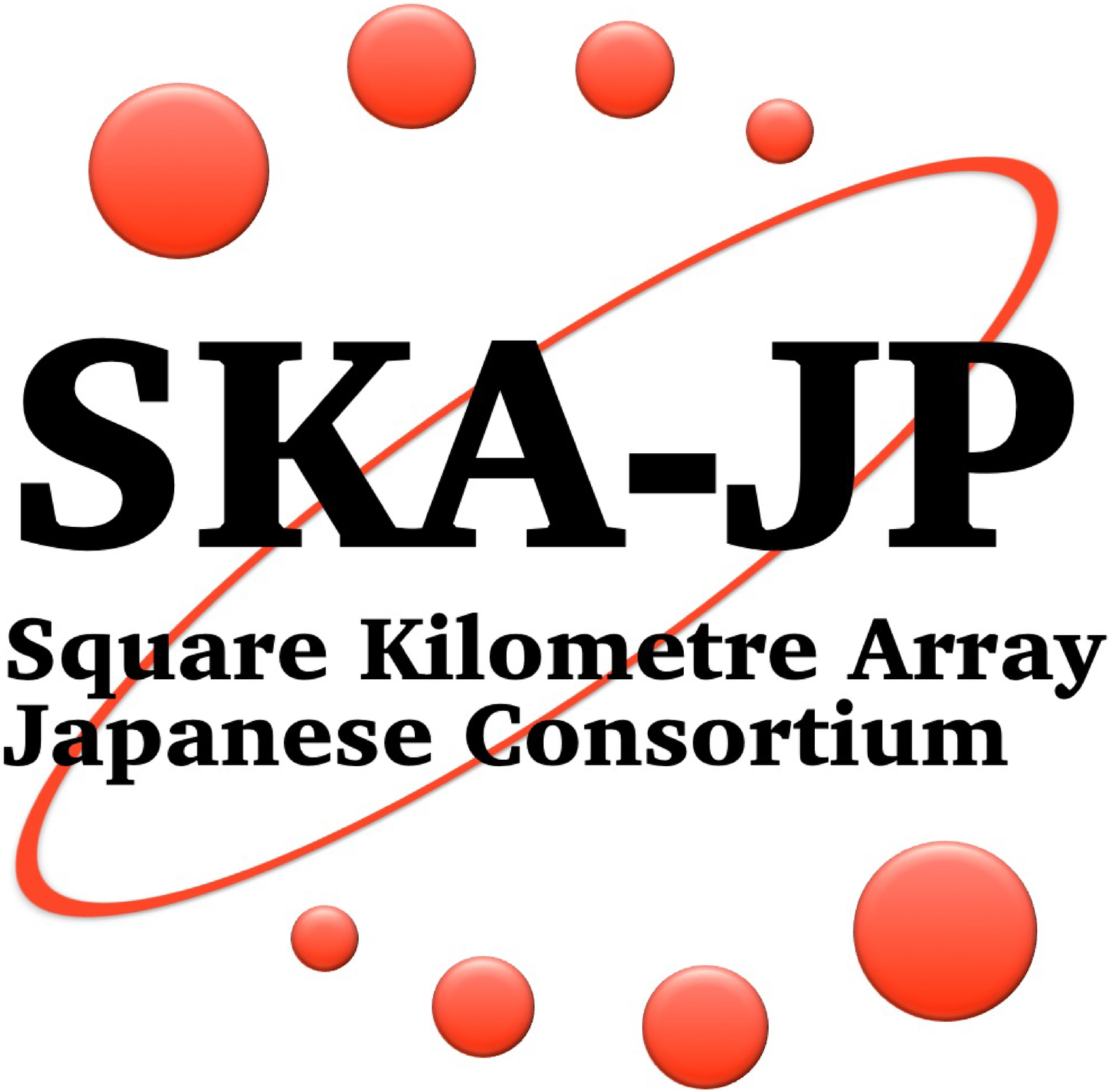}
\vspace{154mm}
\end{figure}

\section{Introduction}

Pulsars are rapidly rotating neutron stars with strong magnetic fields ($\sim 10^8 - 10^{14}~{\rm G}$) and strongly collimated emission. Their masses are typically solar mass and they are considered to be formed as a remnant of supernova explosion. So far more than 2,500 pulsars have been found and there is a wide variety of distinct observational classes such as milli-second pulsars (MSPs) with rotation periods $\sim 10~{\rm msec}$, magnetars with magnetic fields of $\sim 10^{14}~{\rm G}$ and intermittent pulsars which are active only for a few days between periods of a month. Pulsars are extreme objects and their emission mechanism, the properties of magnetosphere and evolution are studied with multi-wavelength observations from radio to gamma-rays.

On the other hand, thanks to the long-term stability of rotation period, pulsars have been used to probe fundamental physics. For example, gravitational waves passing through Earth and pulsars slightly change the arrival times of pulses. Then very precise measurement of the arrival times of pulses from MSPs with very stable rotation periods allows us to detect gravitational waves directly. This method is called pulsar timing array (PTA) and is sensitive to gravitational waves with frequencies $10^{-9}-10^{-7}~{\rm Hz}$, depending on the cadence and period of observation. Thus, PTA plays a part of multi-wavelength gravitational-wave astronomy, together with ground and space laser interferometers which covers much higher frequencies. In fact, the recent direct detection of gravitational waves of $\sim 100~{\rm Hz}$ from a binary black hole merger opened a new window for astronomy \citep{Abbott2016a,Abbott2016b}.

The sources of gravitational waves of the frequencies covered by PTA include supermassive black hole binary and cosmic strings. Currently, three projects, Parkes Pulsar Timing Array (PPTA)\footnote{http://www.atnf.csiro.au/research/pulsar/ppta/}, North American Nanohertz Observatory for Gravitational Waves (NANOGrav)\footnote{http://nanograv.org} and European Pulsar Timing Array (EPTA)\footnote{http://www.epta.eu.org}, are ongoing and beginning to set constraints on stochastic gravitational-wave background from supermassive black hole binaries formed by galaxy collision \citep{Shannon:2015ect}.

Another important application is the test of general relativity in strong gravitational field with neutron star-neutron star and neutron star-black hole binary. Because general relativistic effects appear more strongly for phenomena with larger spacetime curvature, tests with these binaries could put stronger constraints on gravitational theory, compared to conventional weak-field tests. For example, long-term observation of Hulse-Taylor binary PSR B1913+16 revealed its orbital decay, which is in precise agreement with the loss of energy due to gravitational waves predicted by general relativity. This is an indirect detection of gravitational waves. One of the most important target is the binary system of Sgr A${}^*$ and a pulsar orbiting it. In order to perform tests of gravity around the black hole, we need to find a pulsars as close as $\sim 1$mpc ($\sim 0.2$yr in period) from the black hole.

Further, the structure of Galactic ISM and magnetic fields have also been studied with rotation measure and dispersion measure of pulsars. Thus, pulsars are of great interest for a wide community of astronomers and astrophysicists.

The Square Kilometre Array (SKA)\footnote{https://www.skatelescope.org} is a next-generation radio telescope and will bring transformational progress in many fields of astronomy with its overwhelming sensitivity and survey speed. The SKA consists of SKA-low which covers $50-350~{\rm MHz}$ and will be built in Australia and SKA-mid which covers $0.35-10~{\rm GHz}$ and will be built in South Africa. Construction of Phase 1 (SKA1) will begin in 2018 and early science will start from 2020 and it will be scaled up to the full SKA (SKA2) by the late 2020s. The SKA will increase drastically the number of observable pulsar and will revolutionize the study of pulsars themselves and their usage as tools for fundamental physics \citep{Kramer2015}.

The purpose of this article is to show Japanese interests on the pulsar science with the SKA. This article is based on the discussion in the pulsar science working group of SKA-Japan and summarizes our studies and prospects for the SKA pulsar science, focusing on investigation of modified gravity theory with the supermassive black hole in the Galactic Centre, gravitational waves from cosmic strings and binary supermassive black holes, study of the physical state of plasma close to pulsars with giant radio pulses and determination of magnetic field structure of Galaxy with pulsar pairs. In section 2, we summarize the observation strategy of pulsars with the SKA and science objectives developed by the international SKA pulsar science working group. In section 3, we show our studies and interests on the pulsar science with the SKA.

\section{Pulsar Studies with the SKA}

In this section, first we briefly summarize the strategy for pulsar survey and timing observations with the SKA and then introduce the science targets developed by the international SKA pulsar Science Working Group.

\subsection{Observation Strategy}

In order to advance the pulsar sciences, first of all, it is necessary to discover as many pulsars as possible. There are two ways for pulsar search: all-sky survey and targeted searches. Further, for a PTA, high-precision timing will be performed to selected MSPs with high luminosities and stable orbital periods. In this chapter, we summarize the strategy of the SKA for pulsar search and observation. For detail, see \citet{Keane2015}.

\subsubsection{All-sky survey}

To determine the optimum frequency for pulsar survey, the following consideration should be taken into account.
\begin{itemize}
\item Pulsars have steep spectra $S \propto \nu^{-1.6}$ so that they are much brighter at low frequencies.
\item For fixed aperture size or baseline length, the Field of View (FoV) is larger at low frequencies ($\propto \nu^{-2}$).
\item Dispersive delay due to free electrons along the line of sight to the pulsar is smaller at high frequencies ($\propto \nu^{-2}$), although this effect can be removed if the channel bandwidth is enough narrow.
\item Multi-path propagation due to scattering broadens the pulse profile frequency-dependently and is not correctable. This scales as $\propto \nu^{-4}$.
\end{itemize}
Thus, the SKA-low and SKA-mid can perform complementary surveys and it is proposed that the SKA-low surveys high Galactic latitudes ($|b|>5$ degrees) where dispersive delay and scattering are relatively small, and low Galactic latitudes are surveyed by the SKA-mid. It is known that young pulsars ($<10^7$ year) are located mostly in the Galactic plane, while old pulsars like MSPs spread across the entire sky. The surveys with the SKA-low and SKA-mid are complementary in this sense as well.

\begin{figure}[t]
\includegraphics[width=9cm]{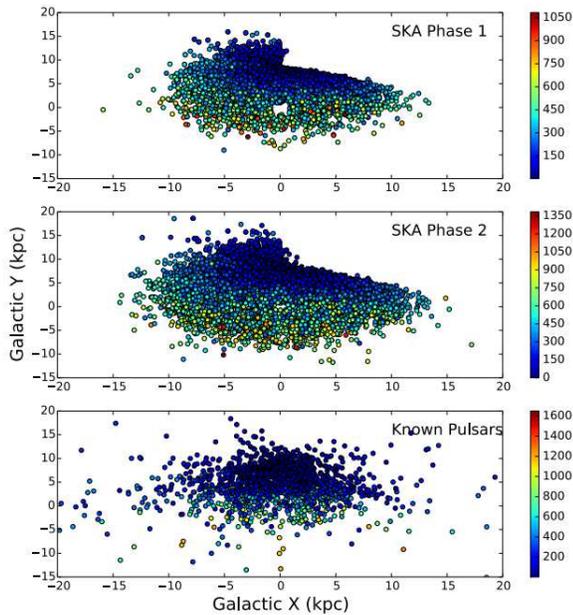}
\caption{Distribution of pulsars expected to be found with the SKA1 and SKA2, and the currently known pulsars, projected onto the Galactic plane \citep{Keane2015}.}
\label{Fig:distribution}
\end{figure}

\citet{Keane2015} performed a series of simulations to estimate the number and distribution of pulsars which the SKA-low and SKA-mid can discover, assuming a fixed integration time of $600~{\rm sec}$ per pointing as described in the baseline design. Further, the SKA-low and SKA-mid are assumed to have 500 and 1500 beams, respectively. They found that SKA1-mid would detect about 9,000 normal pulsars and about 1,400 MSPs, while SKA1-low would detect about 7,000 normal pulsars and about 900 MSPs. On the other hand, with SKA2-low we will find a total of 11,000 pulsars including about 1,500 MSPs, while SKA2-mid will find between 24,000 and 30,000 pulsars, of which between 2,400 and 3,000 will be MSPs depending on the exact improvement in sensitivity. In Fig. \ref{Fig:distribution}, distribution of pulsars expected to be found with the SKA1 and SKA2, and the currently known pulsars, projected onto the Galactic plane.

\subsubsection{Targeted search}

\begin{description}
\item[Galactic Centre \citep{Eatough2015}]
The Galactic Centre is thought to be abundant in pulsars and they could be powerful probes of supermassive black hole, Sgr A${}^*$. The number density of pulsars around Sgr A${}^*$ is estimated to be $\sim 1000~{\rm pc}^3$ \citep{Wharton2012,Chennamangalam2014}, and recently an X-ray and radio loud magneter, PSR J1745-2900, was discovered just $0.1~{\rm pc}$ from Sgr A${}^*$ \citep{Kennea2013,Mori2013}. Because scattering would be serious due to the dense interstellar medium (ISM) for these pulsars, normal pulsars and MSPs need observations with Band 3 ($1.65-3.05~{\rm GHz}$) or 4 ($2.80-5.18~{\rm GHz}$), and 5 ($4.6-13.8~{\rm GHz}$), respectively. Pulsars as close as $\sim 1~{\rm mpc}$ from the supermassive black hole provide unique probes of the spacetime around it and allow us to probe general relativity in strong gravitational field. Especially, Cosmic Censorship Conjecture and no-hair theorem can be tested precisely by measuring its spin and quadrupole moment.
\item[Extragalactic \citep{Keane2015}]
Currently, the only pulsars beyond the disk of the Galaxy are discovered in globular clusters and in the Magellanic Clouds. The SKA will be able to discover pulsars in local group galaxies and will reach even farther galaxies ($\sim 1~{\rm Mpc}$ for giant radio pulses. Extragalactic pulsars allow us to probe intergalactic medium through their dispersion measures, rotation measures and scattering.
\item[globular cluster \citep{Hessels2015}]
The stellar density is so high in globular clusters that pulsars there have much larger chance to experience encounters compared to in lower density environments. Thus, the fraction of MSPs is much larger and exotic binaries such as MSP-MSP or MSP-black hole would be formed. Because globular clusters are compact ($\lesssim$ a few arcmin), a single pointing is enough to cover a whole globular cluster so that pulsar searches in globular clusters are well suitable for the early science. 
\end{description}

\subsubsection{Pulsar timing array}

A PTA consists of a number of MSPs. A stochastic gravitational-wave background causes a change in the time-of-arrival (TOA) of the pulses. In order to extract the gravitational-wave signal from noisy TOA data, the TOAs of each pulsar are correlated to those of other pulsars and the correlated data is analyzed to search for the unique correlation pattern from the quadrupole nature of gravitational waves.  

The noise level of the pulse TOAs is essential for the sensitivity to a gravitational-wave background. The best MSPs with high flux density, narrow pulses and timing stability are selected for the precise timing measurements and observed over many years ($\sim 10~{\rm yrs}$). Usually, only $5-10\%$ of the discovered pulsars have sufficient quality necessary for the PTA. 

The pulsar timing with the SKA is expected to reach the noise level of TOA down to $\sim 100~{\rm nsec}$ by increasing the number of the best MSPs and by reducing the noise levels of plasma propagation effects with the use of multiple telescopes operating in different frequency bands. The SKA1 will monitor 50 MSPs and measure the TOAs once a week for 15 minutes each \citep{Janssen2015}. The observations will be carried out with bands 2, 3 and 4 of SKA-mid while SKA-low will be used for the precise measurement of the variation of dispersion measures.

\subsection{Tests of Gravity Theory}

The tests of gravitational theory is one of high-priority issues in science with SKA. The main project in the international SKA pulsar Science Working Group in this topic is the test of validity of general relativity with binary pulsars based on the parametrized post-Keplerian formalism.

Observation of Type Ia supernovae \citep{Perlmutter:1998np,Riess:1998cb} and other cosmological observations \citep{Adam:2015rua,Ade:2015xua,Ade:2015rim} strongly support the current accelerated expansion of the universe. In the Einstein's general relativity, in order to realize such accelerated expansion of the universe we have to introduce matter fields with negative pressure, called dark energy. On the other hand, there is another possibility that the law of gravity is modified from that of general relativity on large, that is, cosmological scales. The existence of the underlying fundamental gravitational theory beyond general relativity is widely believed for the following reasons. One reason is that, while three among four fundamental forces are described by renormalizable theories, only gravity is not renormalizable. Another reason is that, due to the singularity theorem, general relativity itself predicts the appearance of a singularity where the theory breaks down. The existence of visible singularities, such as big bang singularity or naked singularity might not be resolved without introducing more fundamental theory of gravity. In this situation, modified gravity theory itself draw much attention in various energy scales. 

It is also important to test the theory by observation \citep{1993tegp.book.....W,lrr-2014-4,Koyama:2015vza}
 while the general relativity itself is an elegant theory as a theory of gravity. In fact the general relativity has already passed various observational tests, for example the precession of mercury's perihelion. However, these tests mainly focused on limited aspects of the theory especially in its weak field regime and hence it is also necessary to test other aspects particularly in its strong field regime. Interestingly, observation of pulsars gives us an invaluable opportunity for it \citep{2015aska.confE..42S}. This is quite important because testing gravity in the strong field regime can be regarded as an independent test in the weak field regime. Moreover, we can test the behavior of gravity outside our solar system using pulsars while aforementioned tests are limited to our neighborhood, namely within our solar system.

In the test of gravity with pulsars, our main focus is on ``no-hair theorem of black hole" \citep{Israel:1967wq,Carter:1971zc,Hawking:73,Robinson:1975bv,Hansen:1974zz}. In general relativity in vacuum, a black hole is parametrized only in terms of its mass and angular momentum. This theorem implies that all the multi-pole moments of a black hole are also described by these two parameters. On the other hand, with observation of pulsars, we can independently determine the value of mass, angular momentum and higher multi-pole moments of a black hole. And then it enables us to test the general relativity, especially its aspects in its strong field regime because black hole has strong gravity.

We summarize how various basic parameters of a black hole such as the mass, angular momentum and higher multi-pole moments are determined through the observation of a black hole pulsar binary system. It is known that pulsars emit pulse waves very precisely with certain intervals. However when we observe those pulse waves in reality, those intervals are not constant. Some waves can arrive a little bit earlier and some of others later. This is caused by several reasons. One of the reasons is that if a pulsar rotates around a black hole, the arrival time of pulse waves can be modulated depending on the position of the pulsar where each of the pulse waves is emitted. There are other effects which are associated with astrophysical origins in course of the propagation of pulse wave. Since the general relativistic effects mainly affect the motion of pulsars, we shall focus on it \citep{damour1985general,damour1986general}.

Mathematically, we can parametrize the orbital motion of a pulsar around a black hole in terms of finite number of parameters.
For example, in the Newton's gravity, binary motion or Keplerian motion is characterized only by 6 constant parameters such as orbital period, eccentricity and so on. These 6 parameters are called as Keplerian parameters. However once we take into account general relativistic effects, for example the effect of emission of gravitational waves, the orbital period gets decreased. Related to these facts, we can introduce additional 12 parameters to 
completely parametrize generic motion of a pulsar beyond Keplerian motion from a phenomenological point of view. Those additional parameters are called post-Keplerian parameters. This fact is crucially important since it enables us to parametrize the motion of a pulsar without assuming underlying theory of gravity and hence the result can be applied to any kind of theory of gravity. Then based on this kinematical argument, we can predict the arrival time of pulse waves, which is referred to as ``time of arrival formula", or in short ``TOA formula" \citep{damour1985general,damour1986general}.

On the other hand, once we specify a theory of gravity, those post-Keplerian parameters are determined as a function of Keplerian parameters and the mass of pulsar and black hole. And hence we will be able to observationally determine those post-Keplerian parameters through the information of pulsar arrival time. By choosing one of post-Keplerian parameters we can draw a curve in a plane composed by the mass of pulsar and that of black hole. Then after choosing another two parameters one can further draw two more curves on the same plane. If three curves coincide at one single point, this indicates the theory used is correct. This is how the theory of gravity can be tested with observation of pulsars.

\subsection{Gravitational Waves}

Detection of gravitational waves by PTAs will open an exciting new observational frontier in astronomy and cosmology. The most sensitive frequency of gravitational waves in pulsar timing is determined by the period of observation and is typically $10^{-9}$ to $10^{-7}$Hz. Gravitational-wave sources accessible in this frequency range are supermassive black hole binaries. Although there are many observational evidence for the existence of supermassive black holes even at high redshifts ($z \sim 6$), the detailed mechanism of their formation is far from clear, and is one of the biggest challenges in cosmology. If a PTA succeeds in detecting gravitational waves from supermassive black hole binaries, it would greatly help us to deepen our understanding of the black hole formation history.

The most promising signal in PTA experiments is the stochastic gravitational-wave background from supermassive black hole binaries, which is a superposition of the signals from high-redshift sources to low-redshift sources. The spectrum and amplitude of the background reflect the evolution and population of supermassive black holes and the merger history of galaxies. The current constraints on the strain amplitude by the ongoing PTAs have reached $\sim 10^{-15}$ at $f \sim 0.1~{\rm yr}^{-1}$ and have already excluded some models for the evolution of supermassive black holes \citep{Shannon:2015ect}. Five-year observation of SKA1 PTA will improve the constraint by one order of magnitude and is very likely to detect the gravitational-wave background. If the gravitational waves from nearby sources are strong enough, one may be able even to detect them as individual sources and extract rich astrophysical information by combining data from other observations \citep{Yardley:2010kv,Zhu:2014rta,Arzoumanian:2014gja,Babak:2015lua}. Even in case the signals are not strong enough for the source identification, we may be able to extract information of source distribution by measuring anisotropies in the gravitational wave background \citep{Taylor:2013esa}.

Pulsar timing can also be used to search for gravitational waves from cosmic strings. Cosmic strings are one-dimensional topological defects, which arise naturally in field theories as well as in scenarios of the early Universe based on superstring theory. Strings emit strong gravitational wave bursts from singular structures, called cusps and kinks, and produce a gravitational wave background over a wide range of frequencies, including the sensitivity window of pulsar timing. Currently, there are a few methods to test the existence of cosmic strings such as the gravitational lensing and cosmic microwave background, but the constraint is not very strong since cosmic strings interact with matter only through the gravitational force. Thus, gravitational waves are expected to be a new and powerful observational tool to search for cosmic stings. Their detection or even non-detection of them can be used to constrain cosmic string properties, such as the energy scale of the string and string network evolution, which would provide important insights into particle physics.

\subsection{Galactic ISM and Magnetic Fields}

Radio wave measurements from pulsars make it possible to study the magnetoionic component of the ISM. In this subsection we briefly summarize the current status of research in the free electron distribution, turbulence in the ISM and magnetic fields in the Galaxy following \citet{HAN14}.

\subsubsection{Free electron distribution}

The free electron distribution in the Galaxy is not uniform. A large fraction of the Galactic disc is filled with diffuse ionized gas (DIG) whose electron density is $n_e \sim 10^{-3} - 10^{-2}$ cm$^{-3}$. HII regions are another ionized regions which are distributed following the spiral arms. The range of the electron density in HII regions is large, $n_e \sim 10 - 10^6$ cm$^{-3}$.

\begin{figure}[t]
\vspace{-0cm}
\hspace{0cm}
\includegraphics[width=8cm]{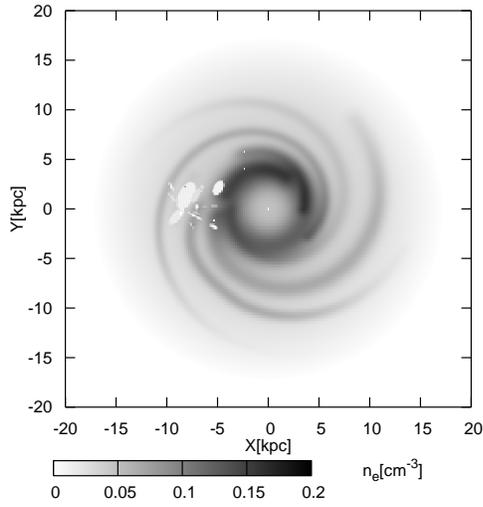}
\vspace{0cm}
\caption{Electron density distribution according to the NE2001 model.}
\label{ne2001z0}
\end{figure}

If distances to pulsars are determined independently, we can obtain the line-of-sight average electron density from dispersion measures (DMs) of pulsars. Since lots of pulsars are distributed in the Galactic disc, it is possible to construct a model of the electron density. \citet{NE2001} developed the detail model (NE2001) which is widely used. Fig.~\ref{ne2001z0} shows the electron density distribution on the Galactic plane($z=0$) according to this model with the original parameters. The Sun is located at $(X,Y)=(-8.5,0)$ in kpc. This model contains the following components: thick and thin discs, spiral arms, a Galactic center, a local ISM, and individual clumps and voids.

We will obtain DMs of a large number of new pulsars which will be found by SKA. Furthermore, distances to the pulsars will be obtained from parallax measurements with SKA-VLBI astrometry \citep{Paragi14}. Consequently, it will be possible to improve the electron density distribution model.

\subsubsection{Turbulence in the ISM}

Radio waves which propagate through the turbulent interstellar plasma are scattered by density irregularities. When we use a thin screen model, an angle of scattering is calculated as $\theta_{\rm scat} \sim (1/2\pi) \sqrt{D/a} r_e \Delta n_e \lambda^2$, where $D$ is the screen thickness, $\Delta n_e$ is the electron density fluctuation, $a$ is a typical dimension of the irregularities, $r_e$ is the classical radius of the electron, and $\lambda$ is the radio wave length \citep{LS1990}.

We can study the interstellar turbulence through observable effects caused by the interstellar scattering. As a result of multipath propagation of the radio waves, observed pulses are broadened particularly at low radio frequencies. By using the thin screen approximation, the pulse broadening time $\tau$ is calculated as $\tau \sim L \theta_{\rm scat}^2/(4c) \propto \nu^{-4}$, where $L$ is the distance to the pulsar. When the density fluctuations follow a Kolmogorov spectrum, the pulse broadening time is proportional to $\nu^{-4.4}$ \citep{LJ1975}. The observed properties of the broadening time are generally consistent with that with the Kolmogorov spectrum \citep{LMGKA2004}. The number of pulsars for which pulse broadening has been measured is $\sim$ 150 as noted in \citet{HAN14}. Because pulse broadening will be measured for a large number of pulsars distributed in the Galaxy with SKA wide-band observations, it will be possible to study the ISM turbulence in different regions in the Galaxy.

Pulsar scintillation is intensity variations caused by density irregularities. A dynamic spectrum and a secondary spectrum of the scintillations are powerful tools for studying the structure of the random irregularities. The dynamic spectrum is a pulse intensity profile in time and frequency space. We can estimate the distribution in size of the irregularities from the dynamic spectrum by using an autocorrelation analysis. The secondary spectrum is the Fourier power spectrum of the dynamic spectrum. The observed secondary spectra show parabolic arcs \citep{ST2006}. We can determine the location of the scattering screen from the curvature of the arcs. Dynamic spectra have been obtained for a few tens of pulsars whose pulse intensity is strong as noted in \citet{HAN14}. It will be possible to obtain dynamic spectra and secondary spectra for lots of pulsars
 through highly sensitive observations by SKA.

\subsubsection{Structure of galactic magnetic fields}

Observations of synchrotron radio emission from spiral galaxies show the existence of magnetic field structures of $\sim \mu$G in strength. Since the energy density of the magnetic fields is comparable to the internal energies of thermal gas and cosmic rays, the roles of the magnetic fields on interstellar gases are important. The magnetic fields are thought to have a large-scale component and a random component. The large-scale field has been modelled by using an axisymmetric (e.g., M31) or bisymmetric configuration (e.g., M51) \citep{RSS1988}. \citet{FBSBH2011} represented the magnetic field in the disc of M51 as a combination of two azimuthal modes $\exp(i m \phi)$ with $m = 0 + 2$, where $\phi$ is the azimuthal angle.

The magnetic field in our Galaxy has been studied by using rotation measures (RMs) of pulsars and extragalactic radio sources. The RM (rad/m$^2$) is given by $\textrm{RM} = 0.81 \int_0^d n_e B_\parallel dl$, where $d$ is the distance to the source in pc, $n_e$ is the number density of the free electrons in cm$^{-3}$, and $B_\parallel$ is the line of sight component of the magnetic fields in $\mu$G.

Previous studies by using the RM distribution along the Galactic plane show that the local field is directed toward the Galactic longitude $l \sim 70^\circ$ - $90^\circ$, and field reversals interior to the solar circle (e.g., \cite{HAN2006}; \cite{VE2011}). The global field strength has been estimated to be $\sim 2~\mu$G, and the amplitude of the random field has been estimated to be $\sim 4-6~\mu$G. \citet{PTKN2011} searched  for the Galactic magnetic field models which reproduce the observed distribution of the RMs of extragalactic radio sources. They found that the magnetic field in the Galactic disc is symmetric with respect to the Galactic plane, and in contrast, the field in the halo region is antisymmetric.

Because of the large amplitude of the random field, distinction between axisymmetric and bisymmetric structures is hardly made. In section 3.4 we will explain the method to estimate the Galactic field structure by using pulsar pairs. Further background of galactic magnetic fields is described in a separate, SKA-JP Magnetism Chapter.

\section{Japanese Science}

In this section, we present the scientific interests of the Japanese pulsar group and our research for the future SKA observations.

\subsection{Supermassive black hole and modified gravity}

Black hole is one of the most fascinating objects predicted in gravitational theories. There are various indirect observational evidences for the existence of black holes in our universe. Environments around such black hole candidates provide us testing sites for new physics under strong gravitational fields. In the international SKA science working group, simulation for gravity test by using the parametrized post-Keplerian formalism has been done \citep{2015aska.confE..42S}. Since the post-Keplerian parameters are purely phenomenological parameters for the binary orbits, even if we detect a deviation from prediction of general relativity in this formalism, the origin or physical interpretation is not immediately clear. Here, we propose two approaches to provide possible origins and physical interpretations for deviation from general relativity. Especially, we focus on the test of gravity around Sgr A* with nearby pulsars which would be discovered by the SKA (see \citet{2015aska.confE..42S} for details of testing gravity with pulsars in the SKA).

\subsubsection{Probe for black hole hair}

As is mentioned in the previous section, SKA enable us to test gravity theories and no hair theorem of black hole by using multipole moment of the black hole. As in general relativity, a modified gravity theory also predict a set of multipole moments of a black hole, which will be in general different from those in general relativity because additional parameters of the theory can enter among such relations. To derive multipole relations we first need to provide rotating black hole solutions in alternative gravitational theories. Apart from general relativity, rotating black hole solutions in modified gravity theories are not so much known (except for few examples e.g. \cite{Yagi:2012ya}). And hence our first task is to find such solutions and theories.

In general, it is very hard to find rotating black hole solutions  in modified gravity theories due to the complexity of field equations. In fact, even in the case of general relativity, it took almost half century for the discovery of the Kerr solution after the advent of general relativity and soon after the discovery of non-rotating black hole solution. Although there is a possibility that the Kerr solution also describes a rotating black hole solution in a specific modified gravity theory, apart from such a special case, we need to develop general formalism to construct solutions for rotating black holes in modified gravity theories. We shall tackle this problem with the aid of perturbative expansion method by which approximated solutions with slow rotation can be generated from a non-rotating solution. It should be noted that non-rotating black hole solutions have been already found in a number of modified gravity theories. Once such slowly rotating solutions are constructed, invaluable indications for rapidly rotating solutions might be given.

To this end, let us first choose a specific model of modified gravity theory to explicitly construct a solution. While there is a bunch of modified gravity theories proposed so far, one of interesting and representative classes of those theories is scalar-tensor theory. The scalar-tensor theory is composed by metric and a scalar field. Due to the presence of a scalar field, the accelerated expansion of the universe can be easily realized. This is why the scalar-tensor theory is often invoked to account for inflation and dark energy. Hence it is interesting if any further observational constraint/verification is given for such a theory. Among them, we are particularly interested in the so-called Horndeski's theory\,(or Generalized Galileon theory), which was originally invented by Horndeski in $1974$ \citep{Horndeski:1974wa,Deffayet:2011gz,Kobayashi:2011nu}.

In Planck units, the action of the Horndeski's theory is given by the following set of equations
\begin{eqnarray}
 S = \sum_{i = 2}^5 \int \dd^4 x \sqrt{- g} {\cal L}_i
\end{eqnarray}
with
\begin{eqnarray}
 {\cal L}_2 &=& {\cal K} (\phi \,, X) \\
 {\cal L}_3 &=& - {\cal G}_3 (\phi \,, X) \Box \phi \\
 {\cal L}_4 &=& {\cal G}_4 (\phi \,, X) R
 - \frac{1}{2} \frac{\partial {\cal G}_4}{\partial X}
 \Bigl[ (\Box \phi)^2
 - (\nabla_\mu \nabla_\nu \phi)^2 \Bigr] \\
 {\cal L}_5 &=& {\cal G}_5 (\phi \,, X) G_{\mu \nu} \nabla^\mu \nabla^\nu \phi \nonumber \\
            & & + \frac{1}{12} \frac{\partial {\cal G}_5}{\partial X}
                \Bigl[ (\Box \phi)^3 - 3 (\Box \phi) (\nabla_\mu \nabla_\nu \phi)^2
                       + 2 (\nabla_\mu \nabla_\nu \phi)^3 \Bigr] 
\end{eqnarray}
where $\phi$ represents a scalar field and $X$ the derivative of the scalar scalar field, that is $X \equiv g^{\mu \nu} \partial_\mu \phi \partial_\nu \phi$. Here ${\cal K} \,, {\cal G}_3 \,, {\cal G}_4$ and ${\cal G}_5$ are arbitrary functions of $\phi$ and $X$ specified by each model. The Horndeski's theory is dubbed as the most general scalar-tensor theory with second order differential equations for metric as well as the scalar field. Then the dynamics of metric and the scalar field is uniquely determined once those initial values and velocities are specified. Since this theory includes various interesting models such as canonical scalar field, ${\cal K}$-essence, $f(R)$ gravity models, it is interesting enough to investigate this particular but still generic class of theory.

In a specific model of the Horndeski's theory, there exist non-rotating black hole solutions \citep{Babichev:2013cya,Sotiriou:2013qea,Minamitsuji:2013ura,Kobayashi:2012kh,Kobayashi:2014wsa,Kobayashi:2014eva,Ogawa:2015pea}. Based on one of those solutions, we have introduced perturbations around a background solution. After consistently solving linearized equations, it can be seen that the solution at the linear order is completely same as that in general relativity \citep{Maselli:2015yva}. And hence there is no deviation from general relativity at this order. However, at the next leading order, there must be difference affecting no-hair relations since the quadrupole moment is a second order quantity. After solving second order equations and analyzing the solution, we will be able to test the theory of gravity, which is our ongoing project right now. 

As for future projects, we plan to apply thus developed techniques to other modified gravity theories. It is also interesting if we can find an exact solution with rapid rotation. Once it is done, we can read multipole relations of black hole spacetime with arbitrary angular momentum (spin).

\subsubsection{Another approach to probe exotic physics around a black hole}

Well-known conventional approaches to probe gravitational theories are the parametrized post-Newtonian (PPN) \citep{Poisson_Will} and the parametrized post-Keplarian (PPK) \citep{damour1986general,1992PhRvD..45.1840D} approaches. Observational constraints on the parameters which characterize deviation from general relativity (GR) in these approaches can be seen in \citep{1993tegp.book.....W,lrr-2014-4}. If the deviation from GR is found around a black hole in the PPN framework, it constraints the alternate gravity theory to some extent. However, in PPN framework, one usually does not take into account effects of possible matter fields around the black hole, so the deviation might come from effects of the matter. This possibility would not be excluded using PPN framework alone. The story of Vulcan might be repeated again for the black hole at the galactic center. Here, based on \citet{Iwata}, we take another phenomenological approach to see how the system deviates from vacuum system in GR.  

We attack this problem in analogy with the dark energy problem in cosmology. It is often assumed that, for simplicity, the dark energy is a perfect fluid with the equation of state $p = w\varepsilon$, where $p$ and $\varepsilon$ are the pressure and the energy density of the dark energy. In order to explain the accelerated expansion of the universe, $w<-1/3$ must be satisfied in GR. However, one may wonder the origin of such exotic matter, and often relies on a modified gravity theory in stead of introducing such exotic matter. In such point of view, indication of existence of the exotic matter may build up our expectation about discovery of new gravitational physics. Following the strategy similar to introducing the dark energy, we consider environment around a black hole based on GR with an exotic matter. Here we note that the exotic matter around a black hole is not necessarily identical to the dark energy in cosmology. We introduce an exotic matter in a purely phenomenological point of view, and do not care about its origin. 

As a first step, let us consider the spherically symmetric system  composed of a black hole and surrounding perfect fluids with the equation of state being $p=w\varepsilon $. Then, we take the post-Newtonian approximation, assuming the perfect fluid can be relativistic, that is, pressure $p$ can make comparable contributions to the curved geometry with the matter density $\varepsilon$. Although, in the usual post-Newtonian expansion, $\varepsilon$ makes Newtonian order contribution at the leading order, we assume that $\varepsilon$ and $p$ make only post-Newtonian contributions. In other word, we treat a sufficiently sparse and relativistic perfect fluid component so that it can be regarded as a small perturbation around the black hole. The main contribution in the curved geometry comes from the Newtonian term associated with the mass of the central black hole. 

In the phenomenological point of view, we may not interested in the global distribution of the perfect fluid, but only local distribution near the region of our interest. When we consider a trapped test-particle motion around this system, we need to specify the fluid distribution at least in a spherical shell domain with the finite width within which the particle moves around. The distribution should be determined by the Hydrostatic equilibrium equation, which is derived from the fluid equation of motion, as follows:
\begin{eqnarray}
\frac{\dd p}{\dd r}
&\simeq& -(1+w) \frac{GM}{c^2} \frac{\varepsilon}{r^2}
\nonumber \\
&\Leftrightarrow& \frac{\dd \varepsilon}{\dd r} \simeq -\frac{1+w}{w}\frac{GM}{c^2}\frac{\varepsilon}{r^2}
\nonumber \\
&\Leftrightarrow& \varepsilon\simeq\varepsilon_R\left[1+\frac{1+w}{w}\frac{GM}{c^2}
\left(\frac{1}{r}-\frac{1}{R}\right)\right], 
\end{eqnarray}
where $M$, $R$ are the black hole mass and the reference radius, and $\varepsilon_R = \varepsilon(R)$. We assume that this distribution is approximately valid only in the spherical shell domain of our interest. 

We introduce the following small parameters:
\begin{equation}
\epsilon=\frac{GM}{c^2R}~,~\alpha=\frac{4\pi R^3\varepsilon_R}{3c^2M}. 
\end{equation}
Then, performing the post-Newtonian iteration and by applying a usual perturbation scheme, we can calculate the value of the pericenter shift. At each order, the contribution to the pericenter shift for 1 period can be estimated as 
\begin{eqnarray}
\Delta \omega_\alpha &\sim& 10^{-2} 
\left(\frac{\varepsilon_R / c^2}{10^{-8}{\rm g/cm^3}}\right)\cr
&&\hspace{1cm}
\times 
\left(\frac{M}{10^6M_\odot}\right)^{-1}
\left(\frac{R}{100~{\rm AU}}\right)^{3}{\rm [rad]}, 
\\
\Delta \omega_{\epsilon\alpha} &\sim& 10^{-6} 
\left(\frac{\varepsilon_R / c^2}{10^{-8}{\rm g/cm^3}}\right)
\left(\frac{R}{100~{\rm AU}}\right)^{2}{\rm [rad]}. 
\end{eqnarray}

It is worthwhile to compare this value with the effect of the black hole spin parameter $\chi := cS/(GM^2)$, where $S$ is the black hole angular momentum. The pericenter shift due to the spin effect $\Delta \omega_{\rm spin}$ can be estimated as 
\begin{equation}
\Delta \omega_{\rm spin}\sim 10^{-6} \chi 
\left(\frac{M}{10^6M_\odot}\right)^{3/2}
\left(\frac{R}{100~{\rm AU}}\right)^{-3/2}~{\rm [rad]}. 
\end{equation}
The spin parameter of Sgr A* could be measured with precision of $\sim 0.1\%$ with the SKA. Therefore, our estimates suggest that, if $\varepsilon_R / c^2 \sim 10^{-12}~{\rm g/cm^3}$, we can measure the value of $\varepsilon_R$ with precision of $\sim 0.1\%$. If $\varepsilon_R / c^2 \sim 10^{-8}~{\rm g/cm^3}$, we can measure the value of $w$ with precision of $\sim 0.1\%$ by using the SKA. 

The observational evidence of $\varepsilon_R<0$ or $w<0$ suggests, with GR to be valid, the existence of an exotic matter, and it might lead up to discovery of new gravitational physics. In order to accurately estimate the measurement precision for $\varepsilon_R$ and $w$, we need to perform detailed simulation as is done in \citet{2012ApJ...747....1L}. Further extensions would be possible, e.g., stationary matter configuration, black hole-pulsar system, non-spherically symmetric configuration, higher-order contribution, and so on. These are our future works as a part of SKA Japan projects.

\subsection{Gravitational waves}

Detection of gravitational waves by PTAs will provide us with rich and unique opportunities to explore new astronomy and cosmology. As the Japanese member of SKA experiment, we will first aim at providing theoretical predictions to lead the pulsar timing data analysis using extant implications for the existence of supermassive black hole binaries obtained from optical observations. Candidates of supermassive black hole binaries have been found in nearby galaxies \citep{Batcheldor:2010vd,Schutz:2015pza} as well as quasars \citep{Liu:2013fca,Graham:2015tba}. Since optical observations determine the positions and masses of the sources, such predictions would help to identify individual sources in PTAs. We will make detailed predictions taking into account the uncertainty in the orbital eccentricity, which affects the amplitude and frequency of the gravitational waves \citep{Yonemaru2016}.

We will also explore the use of information on anisotropy.  Anisotropy in the gravitational wave background may be produced if there is a strong gravitational wave source in a nearby galaxy. A previous work has shown that the anisotropy level depends on the number of nearby sources \citep{Taylor:2013esa}.  We will investigate whether it is possible to estimate the ratio of galaxies having black hole binaries by combining the information of the gravitational-wave background anisotropy and optical observations, and whether it can be connected with the history of black hole formation.

\begin{figure}[t]
\includegraphics[width=8cm]{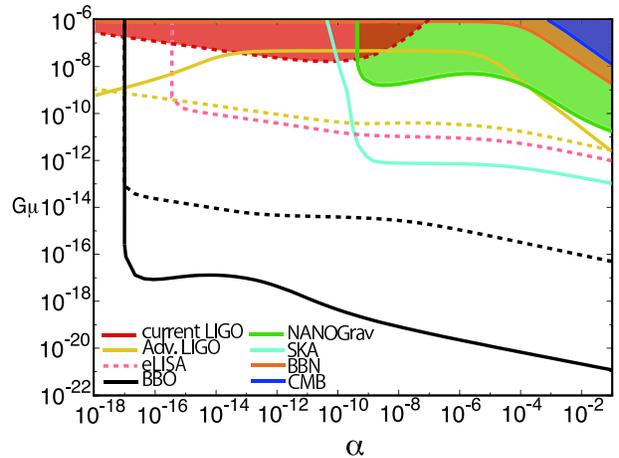}
\caption{Accessible parameter space in the $G\mu -\alpha$ plane (string tension vs. initial loop size) for cosmic strings of a phase-transition origin where the reconnection probability is assumed to be unity \citep{Kuroyanagi:2012wm,Kuroyanagi:2012jf}. The colored regions are excluded by current experiments or cosmological considerations.  The region above the solid or dashed lines is reachable by each future experiment.  Solid lines are for background searches, and dashed lines are for burst searches.}
\label{Fig:cosmicstring}
\end{figure}

We also plan to work on gravitational waves from cosmic string using our knowledge and expertise which we previously developed in our previous works \citep{Kuroyanagi:2012wm,Kuroyanagi:2012jf}.  In the papers, we have estimated the amplitude of the gravitational wave background from cusps on string loops for all the frequencies. Based on these estimates, we have shown the accessible parameter space by several types of future experiments including pulsar timing. As seen in Fig. \ref{Fig:cosmicstring}, the parameter space of $G\mu$ (the string tension or the energy density of the cosmic string) and $\alpha$ (the typical length of string loops normalized by the Hubble length at the time of loop formation) can be explored largely with many types of future gravitatioanl wave experiments.  The SKA (cyan line) would provide the best constraint on $G\mu$ for large values of $\alpha$ among the near-future gravitational wave experiments. We have also shown that different types of gravitational wave experiments can break degeneracies in cosmic string parameters, and provide better constraints than those from each measurement. 

Cusps on string loops are usually considered to be the dominant source of gravitational wave background in pulsar timing sensitivity. However, the main source can be kinks on infinite strings, when the typical initial string loop size in the cosmic string network is small. There are very few estimates for the amplitude of the gravitational wave background from kinks \citep{Kawasaki:2010yi}. We will improve the previous estimate by including the existence of Y-junctions of cosmic super-strings as predicted by super-string theory.  This will enable us to provide more precise predictions on detectability of gravitational waves from cosmic strings.

In addition, we will consider an application of the anisotropy test also to the gravitational wave background of a cosmic string origin. The anisotropy level provides information on the number of cosmic-string bursts, which depends on the evolution of the cosmic string network. While the gravitational wave background consists of bursts from different redshifts in the case of a cusp origin, gravitational waves emitted from a nearby infinite string always dominated in the case of a kink origin.  We will investigate both cases and compare the differences. Using additional information from anisotropy, we will be able to remove the degeneracies of cosmic string parameters, which would give us a better understanding of properties of cosmic strings.

\subsection{Giant radio pulse}

Radio pulses which are stronger than the normal pulses by a factor of $10^3$ - $10^6$ are occasionally observed for some pulsars. They are called Giant Radio Pulses (GRPs) and only about 10 pulsars are reported to have GRPs including the Crab pulsar \cite{Argyle1972} and a mili-second pulsar PSR B1937+21 \citep{Cognard1996}.

The physical mechanism of GRP emission have not been understood, although some theoretical models have been proposed \citep{Petrova2004,Ly07}. For the Crab pulsar, the correlation between GRPs and pulses in other wavelengths were studied. It was reported that optical pulses coincident with GRPs have statistically significant increase in the luminosity of $\sim 3\%$ \citep{Sh03,St13}. The correlation with X-ray emission was studied in \citet{Mi14} and upper bounds on the increase in X-ray luminosity of $O(10)\%$ were obtained. Because the mechanism of high energy emission is relatively well understood \citep{Ta07}, the correlation between the GRPs and high energy emission will enhance the understanding of the emission mechanism of the GRPs.

Strong variability within a GRP of nano-second scale was observed for the Crab pulsar \citep{Ha03,Ha07}. The nano-second variability corresponds to a spatial scale of $O(10)~{\rm cm}$ so that observations of GRPs with high time resolution allows us to probe the physical state of the emission region of the pulsar with this spatial resolution. Significant progress on the understanding of GRPs is expected with this kind of observations and numerical simulations of particle acceleration in the magnetosphere \citep{Ti12}.

The Japanese pulsar group is focusing on the investigation of the density structure of magnetosphere along the path of radio pulses through wideband observations of GRPs. There is an implication that a large amount of high-energy charged particles ($\sim 3 \times 10^{40}~{\rm sec}^{-1}$ for the Crab pulsar) are emitted from the pulsar magnetosphere. However, there has been no clear explanation on the mechanism of the acceleration of the charged particles. We expect the study of the density structure of magnetosphere will provide useful information on the particle acceleration mechanism.

Let us explain the details of the methodology. Due to the dispersive delay in the interstellar plasma, the arrival time of pulses at low frequencies is delayed and the time difference between two frequencies $f_1$ and $f_2$ can be written as \citep{LK04},
\begin{equation}
\label{dispersion}
\Delta t =
\int_0^L \frac{ds}{c \sqrt{1-\left( f_p / f_1 \right)^2}}
- \int_0^L \frac{ds}{c \sqrt{1-\left( f_p / f_2 \right)^2}} 
\end{equation}
where $f_p$ is the plasma frequency,
\begin{equation}
\label{fpe}
f_p
\equiv \sqrt{\frac{n_e q^2}{\pi m_e}}
\sim 9 \sqrt{n_e [\mathrm{cm}^{-3}]} [\mathrm{kHz}]
\end{equation}
and this can be expanded, under a condition $f_p \ll f_{1,2}$, as,
\begin{eqnarray}
\label{dispersion}
\Delta t
&\sim& 4.15 \times 10^{-3} \mathrm{s}
       \left(f_{1,\mathrm{GHz}}^{-2} - f_{2,\mathrm{GHz}}^{-2}\right)
       DM_{\mathrm{pc/cm^{3}}} \nonumber \\
& & + 0.25 \times 10^{-12} \mathrm{s}
      \left(f_{1,\mathrm{GHz}}^{-4}-f_{2,\mathrm{GHz}}^{-4}\right)
      EM_{\mathrm{pc/cm^{6}}}.
\end{eqnarray}
Here,
\begin{equation}
DM \equiv \int_{0}^{L}n_{e} ds, ~~~~~
EM \equiv \int_{0}^{L}n_{e}^{2} ds
\end{equation}
are Dispersion Measure and Emission Measure, respectively.

In the past studies, the second term in Eq. (\ref{dispersion}) was often neglected. However, it is not negligible if there is a localized high-density region along the line of sight, because it is proportional to the square of the plasma density. In fact, \citet{Ku08} reported the time difference which cannot be explained by the first term for the GRPs of the Crab pulsar and obtained an average $EM$ value of $\sim 4 \times 10^6~{\rm pc/cm}^6$ from the simultaneous observations at 44, 63 and 111 MHz. This result implies the possible contribution of the second term in Eq. (\ref{dispersion}) from a high-density region in the Crab Nebula. Considering the variability of the $DM$ value\footnote{``JODRELL BANK CRAB PULSAR MONTHLY EPHEMERIS'', {http://www.jb.man.ac.uk/~pulsar/crab.html}}, the effect of $EM$ may also have a variability which reflects the dynamics of the density structure.

Simultaneous and wide-band observations with the SKA will allow us to measure $DM$ and $EM$ more precisely. Below, we roughly estimate the minimum value of $EM$ which can be measured by the SKA observations. From Eq. (\ref{dispersion}), we see the effect of the $EM$ term is stronger at low frequencies. Therefore, we consider a simple strategy that we determine the $DM$ from the SKA-mid observations of GRPs and then, fixing the $DM$ value, we see if the effect of the $EM$ can be seen in the GRP arrival time for the SKA-low observations. We neglect the effect of the scattering due to inhomogeneous ISM, which is also significant at low frequencies, on the determination of $EM$. Let us denote the observation error in $DM$ as $\delta DM$. Then we define the difference in the time of arrival of pulses at the minimum and maximum frequencies ($f_{\mathrm min}$ and $f_{\mathrm max}$) subtracted by the contribution from $DM$ as $\Delta t'$. From Eq. (\ref{dispersion}), we can write
\begin{eqnarray}
\Delta t'&=& 0.25 \times 10^{-12} \mathrm{s} \left(f_{\mathrm{min},\mathrm{GHz}}^{-4}-f_{\mathrm{max},\mathrm{GHz}}^{-4}\right) EM_{\mathrm{pc/cm^{6}}}
\nonumber \\
&& \pm 4.15 \times 10^{-3} \mathrm{s} \left(f_{\mathrm{min},\mathrm{GHz}}^{-2}-f_{\mathrm{max},\mathrm{GHz}}^{-2}\right)\delta DM_{\mathrm{pc/cm^{3}}},
\label{delay}
\end{eqnarray}
and we obtain,
\begin{eqnarray}
EM_{\mathrm{pc/cm^{6}}}
&=&\frac{\Delta t'}{0.25 \times 10^{-12} \left(f_{\mathrm{min},\mathrm{GHz}}^{-4}-f_{\mathrm{max},\mathrm{GHz}}^{-4}\right)}
\nonumber \\
&& \pm \frac{4.15 \times 10^{-3} \left(f_{\mathrm{min},\mathrm{GHz}}^{-2}-f_{\mathrm{max},\mathrm{GHz}}^{-2}\right)\delta DM_{\mathrm{pc/cm^{3}}}}{0.25 \times 10^{-12} \left(f_{\mathrm{min},\mathrm{GHz}}^{-4}-f_{\mathrm{max},\mathrm{GHz}}^{-4}\right)}.
\nonumber \\
\label{eq:EM}
\end{eqnarray}
By evaluating the second term of Eq. (\ref{eq:EM}), we estimate the minimum measurable value of EM with SKA. Adopting $f_{\rm min} = 0.05~{\rm GHz}$, $f_{\rm max} = 0.35~{\rm GHz}$ and $\delta DM = 0.001~{\rm pc/cm}^3$,
\begin{equation}
EM_{\mathrm{min, SKA}} \simeq 4.1 \times 10^4~[\mathrm{pc/cm^6}],
\end{equation}
while the same argument for the observation in \citet{Ku08} leads to
\begin{equation}
EM_{\mathrm{min, Kuz'min}}
\simeq 1.4 \times 10^5~[\mathrm{pc/cm^6}].
\end{equation}
Thus, the SKA observations can constrain $EM$ more accurately by a factor of 3 compared to the current observation even if we assume the same error in $DM$. The effect of $EM$ has not been studied for GRPs of other pulsars. In particular, since intrinsic duration of GRPs of milli-second pulsars are short \citep{So04, Bilous2015}, the SKA-mid will be necessary to study their $DM$ and $EM$.

For observations of GRPs, we can expect much larger signal-to-noise ratio than normal pulses and a single GRP can be observed without integration. Thus, we can measure $DM$ and $EM$ for each GRP and their variability of min-hour time scales can be probed. These information can put constraints on the size of high-density region and will provide us a hint to understand the physical state of the magnetosphere. Further, as the density of the dense region increases, the plasma frequency $f_p$ approaches to the observation frequency $f$ and the group velocity of the radio wave approaches to zero. In this case, a cutoff would be observed at the low frequency side.

As we saw above, wide-band observations with the SKA will allow us to investigate the physical structure of pulsar magnetosphere with an unprecedented depth. Correlation of this with other quantities such as the pulsar age and spin-down luminosity will also be of interest.

\subsection{Galactic magnetic field analysis using pulsar pairs}

In this section we explain the method to estimate the Galactic magnetic field structure by using pulsar pairs.

When we use a pair of two pulsars which are seen in almost the same direction. The electron density weighted magnetic field $B_\parallel$ ($\mu$G) along the line of sight in the region between the pair is obtained in terms of differences between their RMs ($\Delta RM$) and DMs ($\Delta DM$),
\begin{equation}
\label{bpara}
B_\parallel = \frac{\Delta RM}{0.81 \Delta DM}
\end{equation}
\citep{OS1993}.
Since many subspaces between the paired pulsars are distributed in the Galaxy, we may obtain a three dimensional structure of the large scale field.

When a pair has no angular separation on the sky $\Delta \theta = 0$, Eq.(\ref{bpara}) represents the correct field strength of the subspace between the paired pulsars, because two radio waves emitted by the paired pulsars propagate through common interstellar space between the closer pulsar and the observer. Since observed pair pulsars have finite angular separation $\Delta \theta$, $\Delta RM$ and $\Delta DM$ do not represent exact RM and DM between the paired pulsars. Therefore, $B_\parallel$ calculated by eq. (\ref{bpara}) deviates from exact $B_\parallel$ as discussed in \citet{OS1993}. In order to reduce this effect, \citet{HAN2004} used the following criterion for $\Delta \theta$,
\begin{equation}
\label{han}
D_1 \sin(\Delta \theta) < 0.1 S
\end{equation}
where $D_1$ is the distance between the Sun and the pulsar which is assumed to be closer to the Sun and $S$ is the distance between the paired pulsars \citep{HAN2004}.

\begin{figure}[t]
\vspace{0cm}
\hspace{0cm}
\includegraphics[width=7cm]{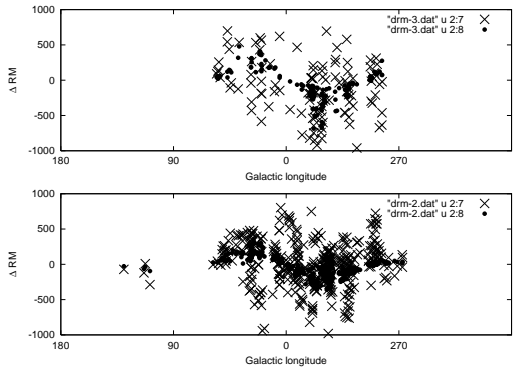}
\vspace{0.5cm}
\caption{$\Delta RM$ distribution along the Galactic longitude. The cross symbols represent $\Delta RM$s for the observed pulsar pairs. The filled circles represent the calculated $\Delta RM$s with the ring field model. The upper panel is for the case of 121 pairs which are located in the region with the distance  4 kpc $< d <$ 8 kpc, and the lower panel is for the case of 404 pairs with 2 kpc $< d <$ 6 kpc.}
\label{ringdrml}
\end{figure}

We show $\Delta RM$ distribution along the Galactic longitude in Fig. \ref{ringdrml}. The cross symbols represent $\Delta RM$s for the observed pulsar pairs taken from the pulsars located at the Galactic latitude $|b| < 5^\circ$ and the distance $d < 12$ kpc. We have used the condition eq.(\ref{han}) for $\Delta \theta$s. The pulsar data are drawn from the Australia Telescope National Facility (ATNF) Pulsar Catalogue \citep{Ma05}. The upper panel is for the case of 121 pairs which are located in the region with the distance  4 kpc $< d <$ 8 kpc, and the lower panel is for the case of 404 pairs with 2 kpc $< d <$ 6 kpc.

We will compare the observed $\Delta RM$ distribution with that calculated by assuming the global field distribution and the electron density distribution in the Galactic disc. We calculate RMs for virtual pulsars, which have the same positions ($l$, $b$, $d$) as the observed pulsars, and calculate $\Delta RM$s for the same pulsar pairs as the observed pairs. We used the NE2001 model as the electron density distribution \citep{NE2001}. The distance between the Sun and the Galactic center is assumed to be 8.5 kpc.

The filled circles in Fig. \ref{ringdrml} represent the calculated $\Delta RM$s in the case when we assume the ring field in the Galactic disc. We set the radial and the vertical components $B_r = 0$, $B_z=0$. The azimuthal component is given by
\begin{equation}
B_\phi = B_0 \sin(\pi R/R_{\rm rev}), 
\end{equation}
where $R$ is the radial distance, and $R_{\rm rev}$ is the radial distance at which the field reversal occurs. We adopt $R_{\rm rev} = 8$ kpc and $B_0 = 2 \mu$G.

\begin{figure}[t]
\vspace{0cm}
\hspace{0cm}
\includegraphics[width=7cm]{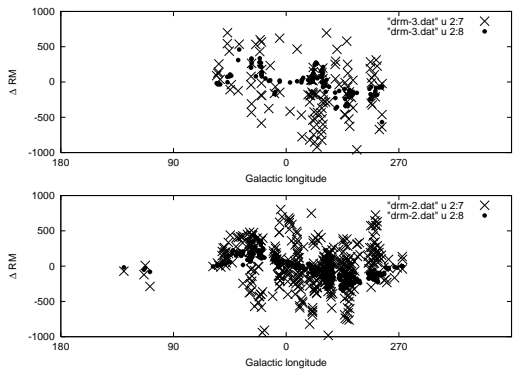}
\vspace{0.5cm}
\caption{The same as Fig. \ref{ringdrml}, but for the BSS field model.}
\label{bssdrml}
\end{figure}

Fig. \ref{bssdrml} shows the case when we assume the bisymmetric spiral (BSS) field model. The radial and azimuthal components are given by
\begin{equation}
B_r = B_0 \sin(\chi) \cos \left[ \frac{1}{\tan (\chi)} 
                                \ln \left ( \frac{R}{R_{\rm sun}} - \phi + \phi_0 \right )
                         \right],
\end{equation}
\begin{equation}
B_\phi = B_0 \cos(\chi) \cos \left [ \frac{1}{\tan (\chi)} 
                                \ln \left ( \frac{R}{R_{\rm sun}} - \phi + \phi_0 \right )
                         \right ],
\end{equation}
where $\chi$ is the pitch angle and the $\phi_0$ is an arbitrary phase. The pitch angle is $\chi = 15^\circ$ in this calculation. The arbitrary phase $\phi_0$ is taken as $270^\circ$ so that the local field is directed toward $l \sim 70^\circ - 90^\circ$.

The lower panels of Figs.\ref{ringdrml} and \ref{bssdrml} (the case with the distance 2 kpc $< d <$ 6 kpc) show the similar variation of $\Delta RM$ along the Galactic longitude between the ring field model and the BSS model, because the local field direction is toward $l \sim 70^\circ - 90^\circ$ in both field models. However, we can find a difference between two models in the upper panels of Figs.\ref{ringdrml} and \ref{bssdrml} (the case with the distance 4 kpc $< d < 8$ kpc). The $\Delta RM$ distribution with the ring field model is antisymmetric with respect to $l = 0$, because the magnetic fields in $R < 8$ kpc are counterclockwise when we view from the north Galactic pole. In contrast, the $\Delta RM$ distribution with the BSS field model is symmetric with respect to $l \sim 0$, because this BSS fields in $R ~< 5$ kpc are counterclockwise in $l > 0$ and clockwise in $l < 0$. 

The calculated $\Delta RM$ distributions vary depending on the magnetic field models. However, because the observed $\Delta RM$s have large scatter, distinction between the ring and BSS models is not possible. If $\Delta RM$s are obtained from a large number of pulsar pairs, we can determine the global field structure from a profile of $\Delta RM$ distribution, but also we can estimate the random field strength from dispersion of $\Delta RM$ distribution.

When we use pulsar pairs, it is possible to study $\Delta RM$ distributions of different samples with different distances as shown in Figs.\ref{ringdrml} and \ref{bssdrml}. This is one of advantages of this method. We will discover more than 10,000 pulsars through SKA all-sky survey. A large data set of RMs and DMs provided by SKA will bring significant progress in the study of the Galactic magnetic field with the analysis using pulsar pairs. Especially, new findings on turbulent fields at short wavelength will come out.

\section{Summary}

In this article, we gave an overview of Japanese scientific interests on SKA pulsar research, based on the discussion in the SKA-Japan pulsar Science Working Group. The unprecedented sensitivity and survey speed of the SKA will enable transformational pulsar sciences which have a significant impact on both astrophysics and fundamental physics. In particular, we focused on probing modified gravity through supermassive black hole at Galactic Centre, gravitational waves from cosmic strings and their anisotropy, giant radio pulses to probe physical state of pulsar magnetosphere and Galactic magnetic fields through pulsar pairs. Beside these topics, our pulsar group is considering a possible collaboration with the VLBI group, especially pulsar astrometry to enhance the sensitivity of pulsar timing array on the gravitational-wave detection.

\begin{ack}
The authors are grateful to the international SKA pulsar Science Working Group for providing us opportunities of open discussion and cooperation. The work of K.T. is supported by Grand-in-Aid from the Ministry of Education, Culture, Sports, and Science and Technology (MEXT) of Japan, No. 24340048, No. 26610048 and No. 15H05896. The work of A.N. was supported in part by the JSPS Research Fellowship for Young Scientists No. 263409.
\end{ack}


\end{document}